\documentclass[twocolumn,tighten,times]{aastex701}
\usepackage{xcolor}
\usepackage{txfonts}

\newcommand{\hcop}{HCO$^+$~}

\newcommand{\msun}{M$_{\odot}$~}

\newcommand{\mujyb}{$\mu$Jy~beam$^{-1}$}
\newcommand{\kms}{km~s$^{-1}$}

\newcommand{\revision}[1]{#1}

\begin{document}

%\title{Arp220 Polarization}
\title{The magnetic fields of the dusty nuclei and molecular outflows of Arp\,220}

\author[orcid=0000-0001-5357-6538]{Enrique Lopez-Rodriguez}
\affiliation{Department of Physics \& Astronomy, University of South Carolina, Columbia, SC 29208, USA}
%\affiliation{Kavli Institute for Particle Astrophysics \& Cosmology (KIPAC), Stanford University, Stanford, CA 94305, USA}
\email[show]{elopezrodriguez@sc.edu}

\author[0000-0002-3829-5591]{Josep M. Girart}
\affiliation{Institut de Ci\`encies de l'Espai (ICE-CSIC), Can Magrans s/n, E-08193 Cerdanyola del Vall\`es, Catalonia, Spain}
\affiliation{Institut d’Estudis Espacials de Catalunya (IEEC), Campus del Baix Llobregat—UPC, Esteve Terradas 1, E-08860 Castelldefels, Catalonia, Spain}
\email[]{girart@ieec.cat}  

\author[0000-0001-5654-0266]{Miguel P\'erez-Torres}
\affiliation{Instituto de Astrof\'isica de Andaluc\'ia, Consejo Superior de Investigaciones Cient\'ificas (CSIC), Glorieta de la Astronom\'ia s/n, E-18008, Granada, Spain}
\affiliation{School of Sciences, European University Cyprus, Diogenes street, Engomi, 1516 Nicosia, Cyprus}
\email[]{torres@iaa.csic.es}

%Mar Mezcua tiene las mismas afiliaciones que Josep Miquel, Ruben y Chema. He puesto las afiliaciones de Miguel, Antxon y Gemma de sus ultimos articulos. Todos tendremos que revisar estas afiliaciones

\author[0000-0003-4440-259X]{Mar Mezcua}
\affiliation{Institut de Ci\`encies de l'Espai (ICE-CSIC), Can Magrans s/n, E-08193 Cerdanyola del Vall\`es, Catalonia, Spain}
\affiliation{Institut d’Estudis Espacials de Catalunya (IEEC), Campus del Baix Llobregat—UPC, Esteve Terradas 1, E-08860 Castelldefels, Catalonia, Spain}
\email[]{}  

\author[0000-0002-2189-6278]{Gemma Busquet}
\affiliation{Departament de F\'{\i}sica Qu\`antica i Astrof\'{\i}sica (FQA), Universitat de Barcelona, Mart\'{\i} i Franqu\`es 1, E-08028  Barcelona, Catalonia, Spain}
\affiliation{Institut de Ci\`encies del Cosmos (ICCUB), Universitat de Barcelona, Mart\'{\i} i Franqu\`es 1, E-08028 Barcelona,  Catalonia, Spain}
\affiliation{Institut d’Estudis Espacials de Catalunya (IEEC), Campus del Baix Llobregat—UPC, Esteve Terradas 1, E-08860 Castelldefels, Catalonia, Spain}
\email[]{}

\author[orcid=0000-0002-7758-8717]{Rub\'en Herrero-Illana}
\affiliation{Institut de Ci\`encies de l'Espai (ICE-CSIC), Can Magrans s/n, E-08193 Cerdanyola del Vall\`es, Catalonia, Spain}
\affiliation{Institut d’Estudis Espacials de Catalunya (IEEC), Campus del Baix Llobregat—UPC, Esteve Terradas 1, E-08860 Castelldefels, Catalonia, Spain}
\email[]{}

\author[0000-0002-9371-1033]{Antxon Alberdi}
\affiliation{Instituto de Astrof\'isica de Andaluc\'ia, Consejo Superior de Investigaciones Cient\'ificas (CSIC), Glorieta de la Astronom\'ia s/n, E-18008, Granada, Spain}
\email[]{}

\author[0000-0002-6896-6085]{Jos\'e M. Torrelles}
\affiliation{Institut de Ci\`encies de l'Espai (ICE-CSIC), Can Magrans s/n, E-08193 Cerdanyola del Vall\`es, Catalonia, Spain}
\affiliation{Institut d’Estudis Espacials de Catalunya (IEEC), Campus del Baix Llobregat—UPC, Esteve Terradas 1, E-08860 Castelldefels, Catalonia, Spain}
\email[]{torrelle@ieec.cat}  

%\author{Add your name}
%\affiliation{Add your institution}
%\email[]{Add your email}  

%\author[orcid=0000-0000-0000-0002,gname=Bosque, sname='Sur America']{Forrest Sur Am\'{e}rica} 
%\altaffiliation{Las Campanas Observatory}
%\affiliation{Universidad de Chile, Department of Astronomy}
%\email{fakeemail2@google.com}

\begin{abstract}

Galaxy mergers trigger starburst activity and galactic outflows that enrich the circumgalactic medium, profoundly impacting galaxy evolution. These phenomena are intrinsically linked to the physical conditions of the medium, which is permeated by magnetic (B) fields affecting its transport and dynamics. Here, we spatially resolve, $0\farcs24$\,(96\,pc), the B-fields in the dusty and molecular outflows of  Arp\,220, the closest ($78$\,Mpc) Ultra-Luminous Infrared Galaxy hosting two interacting nuclei, denoted as East and West. We perform ALMA $870\,\mu$m dust continuum polarization and CO(3-2) emission line polarization, and report the first detection of CO(3-2) emission line polarization through the Goldreich-Kylafis effect in an outflow. Dust polarization shows that Arp\,220\,E has a spiral-like B-field on the disk with a linear polarization fraction of $0.4\pm0.1$\% that may produce the detected circular polarization passing through foreground aligned dust grains. Arp\,220\,W reveals a B-field parallel to the red- and blueshifted outflows in both the dust and emission line polarization maps. The outflows show a dust polarization of $0.2$\%, while the CO(3-2) emission line polarization is $1-2$\% at $4-6\sigma$ significance across independent velocity channels. A highly polarized ($3-5$\%) dusty bridge has a B-field orientation of $\sim110^{\circ}$ connecting both nuclei. Mean B-field strengths of $1.1$\,mG and $9.5$\,mG for the blue- and redshifted outflows, respectively, are estimated. These strong B-fields are attributed to amplification by compression in nuclear clouds and supernova remnants. This amplified B-field is likely sustained by the turbulent kinetic energy in the outflow and may be critical in directing the transport of metals and cosmic rays into the circumgalactic medium.

\end{abstract}



\section{Introduction}\label{sec:INT}

%%%%%%%%%%%%%%
\begin{figure*}[ht!]
\includegraphics[width=\textwidth]{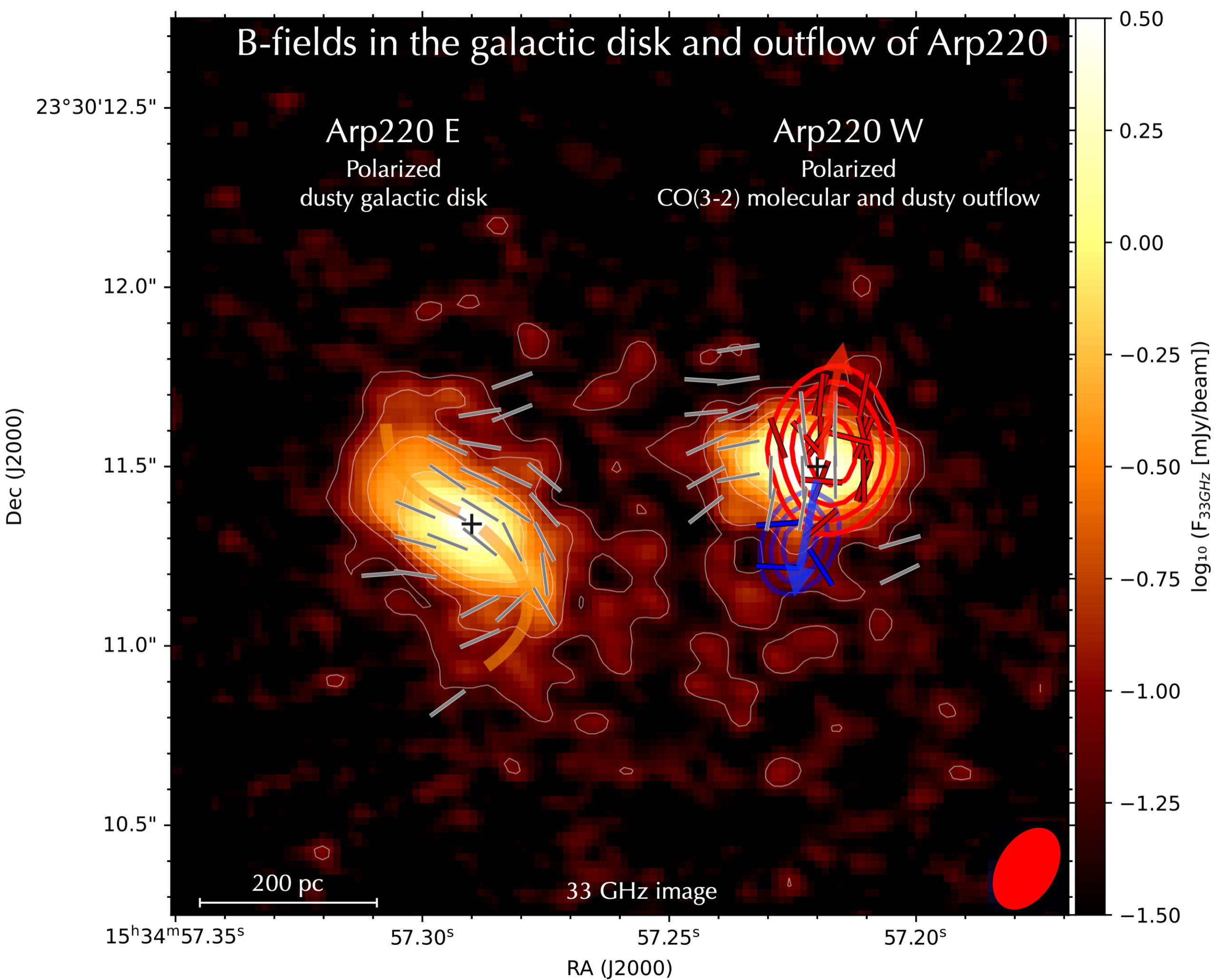}
\caption{The dusty and CO(3-2) molecular B-field orientations in Arp\,220. The disks of Arp\,220\,E and Arp\,220\,W observed at 33~GHz radio wavelengths \citep[tracing syncrotron emission;][]{BM2015} are shown with the overlaid of our measured B-field orientations. These B-field orientations are derived from the linear polarization of the $870\,\mu$m dust continuum (grey lines; see Section~\ref{sec:DustPol} and Fig.~\ref{fig:fig2}) and the CO(3-2) molecular outflow (red and blue lines; see Section~\ref{sec:COPol} and Fig.~\ref{fig:fig3}) observed with ALMA. For reference, the figure also indicates a spiral B-field (orange line), the bipolar molecular outflow direction in Arp\,220\,W (blue and red arrows), the beam size ($0.24\arcsec\times0.16\arcsec$; $96\times60$ pc$^{2}$ oriented at $-31^{\circ}$) of our ALMA continuum observations (red ellipse), the central position of both cores (black crosses), and a physical scale of 200~pc. 
 \label{fig:fig0}}
\end{figure*}
%%%%%%%%%%%%%%

Galaxy evolution is fundamentally driven by hierarchical merging, where these major events perturb the gravitational and gaseous environment both within and around galaxies  
\citep{Riess1998,Keres2009}. Under this scenario, the gravitational torques induced by galaxy interaction effectively destabilize gas disks, causing the gas to lose angular momentum and increasing gas turbulence and funneling it inward \citep{Hernquist1989,BH1996,Hopkins2008}. This gas infall increases turbulence from kiloparsec scales down to the central nuclear regions. The inflowing gas and resulting mergers can trigger intense starburst activity \citep{Smail1997,EC2003,Tacconi2010}, which, in turn, generates powerful galactic outflows that enrich the surrounding circumgalactic medium (CGM) with freshly produced metals, gas, and dust. The physics governing these phenomena is deeply rooted in the properties of the interstellar medium (ISM), which is permeated by magnetic fields (B-fields) in close equipartition with the gas kinetic energy. B-fields are actively amplified by turbulent dynamics, primary driven by supernova explosions \citep{Schober2013}. These amplified B-fields are central to the dynamics of galactic feedback, as they significantly affect the transport and containment of ionized gas, dust, metals, and cosmic rays moving outwards to the CGM via galactic outflows \citep{LR2023}. One notable confirmation from high-resolution magnetohydrodynamical (MHD) cosmological simulations has been that weak B-fields in galaxies are amplified by turbulence in the ISM \citep{Pakmor2014,Rieder2016,MA2022}. This B-field amplification becomes progressively more directly driven by supernova feedback at lower redshifts \citep{MA2018}. These findings suggest a strong correlation between B-field amplification, stellar feedback (manifested as galactic winds), and the injection of turbulence by star formation. Consequently, star-forming galaxies are positioned as one of the most significant contributors to the process of B-field amplification, with profound consequences for galactic feedback.

Indeed, the strongest B-fields, typically ranging from $50$ to $300\,\mu$G, are found in the cores of nearby starburst galaxies, where their ordered B-fields trace the orientation of the galactic outflows \citep{LR2021,LR2023,Pattle2021,Borlaff2023,Belfiori2025}. These findings were obtained using far-infrared (FIR; $53-214\,\mu$m) and sub-mm ($\sim850\,\mu$m) polarimetric observations, which are sensitive to magnetically aligned dust grains in the cold and dense ISM--the region where the bulk of a galaxy's mass and star formation activity is concentrated. FIR and sub-mm thermal polarized emission observations primarily trace the density-weighted B-field integrated over the telescope's beam \citep{LR2022,MA2024}. Following these studies, and with the specific goal of quantifying the B-fields within a merging system exhibiting outflows, we have targeted Arp\,220. Located at $78$ Mpc ($400$\,pc $= 1$\arcsec), Arp\,220 is the closest Ultra-Luminous Infrared Galaxy \citep[ULIRG; $\log(L_{\rm{IR}}/{\rm L_{\odot}}) = 12.19$;][]{PS2021}. Figure \ref{fig:fig0} summarizes the system and our findings. Arp\,220 is a late-stage merger system with a total stellar mass of $\log (M_{\star}/{\rm M_{\odot}}) = 10.81$ \citep{U2012} containing two counter-rotating highly obscured and compact, $\sim150$ pc, nuclei (Arp\,220\,E and Arp\,220\,W) separated by $\sim370$ pc. Both nuclei are sites of strong star formation, driving rates exceeding $200$ \msun${\rm yr}^{-1}$ \citep{BM2015}. A particularly prominent characteristic is the detection of high-velocity molecular outflows ($>400$\,km~s$^{-1}$ relative to the systemic velocity of $5515.6$\,km~s$^{-1}$) emanating from both the West and East nuclei \citep[e.g.,][]{Baan1989,Sakamoto1999,Sakamoto2009,Sakamoto2017,Rangwala2011,Scoville2017,Martin2016,BM2018,Wheeler2020}. Given its extremely dusty and molecular nature, Arp\,220 serves as an excellent system to quantify the effect of B-fields on both its ISM and outflows.

Estimates for the B-field strength in Arp\,220 exhibit a substantial variance, spanning two orders of magnitude  \citep[$0.25-27.5$\,mG;][]{Thompson2006,McBride2014,McBride2015}. The significant variation is primarily a result of the different phases of the ISM probed by the respective observing techniques. For instance, maser clouds within the nuclear regions yield B-fields strengths in the mG range, whereas sub-mG fields are detected in the more diffuse ISM of the system. In all cases, these B-fields measurements were obtained via radio polarimetric observations sensitive to the diffuse and hot ISM, consequently missing the B-field structure within the cold and dense ISM, star-forming regions, and outflows of Arp\,220. Only a limited number of previous studies have attempted to investigate the B-field properties using the signature of magnetically aligned dust grains in Arp\,220 \citep{JK1989, Seiffert2007, Clements2025}. These previous works utilized unresolved polarimetric observations and consistently reported a polarization fraction in the range of $\sim1-2$\% and a plane-of-the-sky B-field orientation at a position angle (PA) of $\sim80-100^{\circ}$, which effectively traces the dust ridge (`overlap') connecting the two nuclei.

In this work, we investigate the B-field properties and their role in the disk and outflows of Arp\,220. We perform polarimetric observations at $870\,\mu$m  with an angular resolution of $\sim0\farcs24$ ($\sim96$\,pc at the Arp\,220 distance) using the Atacama Large Millimeter/submillimeter Array (ALMA). Our goals are to spatially resolve the merger and simultaneously capture full polarization in both dust continuum and CO(3-2) molecular line emission.  Aspherical dust grains aligned with the B-fields produce linear polarization \citep{Andersson2015, Giang2023}, which allows to trace the B-fields in the plane of the sky \citep[e.g.,][]{Girart2006, Huang2024}. Furthermore, molecular lines can also become linearly polarized in the presence of a B-field through the Goldreich-Kylafis effects \citep[GK; e.g.,][]{G-K1982, Ching2016, Lee2018}.  \revision{The GK effect occurs due to the different absorption probabilities of the sublevel transitions $\Delta m \pm1$ ($\sigma$) and $\Delta m =0$ ($\pi$), when resonant radiation is anisotropic. When any of these transtions is favored, it causes an overpopulation of some magnetic sublevel with respect to the other, where $\sigma$ produces polarization angles perpendicular to the B-field and $\pi$ parallel to the B-field.} The combination of these two effects—dust polarization and molecular line polarization—can be utilized to provide a comprehensive description of the B-field properties within the dusty and molecular medium \citep{Girart1999, Lai2003, Cortes2021}. Our findings provide constraints on the role of B-fields in galactic feedback mechanisms. We describe the observations and data reduction in Section \ref{sec:OBS}. We present the results for the dust polarization, emission line polarization, and circular polarization in Section \ref{sec:RES}. The discussion and conclusions are presented in Section \ref{sec:DIS}.

%~\\

%\textbf{MPT: Habría que revisar brevemente estudios previos de polarización de polvo extragaláctico y enfatizar que no se ha observado polarización en líneas de emisión.}

\section{Observations and Data Reduction} \label{sec:OBS}

%From Chema: Josep Miquel, Enrique he retocado muy ligeramente el siguiente párrafo. Podéis comprobar que no haya metido la pata?

The ALMA full polarization observations (project ID: 2023.1.00044.S; PI: Josep Miquel Girart) were carried out in June 2024 in Band 7 ($345.79$\,GHz; $870\,\mu$m) with the C43-5 configuration. The correlator was set to have four spectral windows with a bandwidth of 1.875 MHz. Two of them, centered at 343.9 and 358.1 GHz, were dedicated to the continuum. The other two windows were centered at the rest frequency of the CO(3--2) ($345.8$ GHz) and \hcop(4--3) ($356.7$ GHz) lines. These line windows provided a spectral resolution of $4.2$~\kms. Here, we present the polarization observations of CO(3--2) and dust continuum. Self-calibration was performed independently for the CO(3--2) line data and the two continuum windows. 
The self-calibration allowed to increase the dynamic range from 360 to 2100 in the continuum, and from 90 to 360 in the CO(3--2) images. The CO(3--2) channel maps were created by averaging four channels and using natural weighting, resulting in a synthesized Full-Width Half-Maximum (FWHM) size of $0\farcs34\times0\farcs24$ and a PA of $-35$\degr. The resulting spectral resolution of the channel maps is 13.5~\kms. The median $rms$ noise across all channels for all Stokes parameters is $\sigma_{\rm CO} = 0.27$\,mJy beam$^{-1}$. The full polarization continuum images were obtained using a Briggs's robust weight of 0.5 \citep{Briggs1995}, which yielded a beam size of $0\farcs24\times0\farcs16$ at a PA of $-31$\degr. The dust continuum image (Stokes $I$) has a sensitivity of $\sigma_{\rm{I}} = 99$~\mujyb, while the polarized signal images (Stokes $Q$ and $U$) have a sensitivity of $\sigma_{\rm{QU}} = 18.9$~\mujyb.

For the polarimetric analysis, we only use polarization measurements with $PI/\sigma_{\rm PI} \ge 4$, where $PI$ is the polarized intensity and $\sigma_{\rm{PI}}$ is its uncertainty. For the CO(3-2) line emission polarization analysis, we apply the additional constraint $I_{\rm{CO}}/\sigma_{\rm{I}_{\rm{CO}}} \ge 140$. This stringent condition ensures that each polarization \revision{fraction} measurement has an uncertainty of $<1\%$ based on white-noise polarimetric statistics, and the total intensity is spatially located specifically at the cores of Arp\,220. The peak positions for both merging nuclei, as determined from the dust continuum image, are (RA, DEC)$_{\rm{E}}$ = (15:34:57.29, 23:30:11.34) and (RA, DEC)$_{\rm{W}}$ = (15:34:57.22, 23:30:11.50). For the photometric analysis, we adopt the 10\% flux calibration uncertainty in Band 7 as provided by ALMA\footnote{ALMA technical book: \url{https://almascience.nrao.edu/proposing/technical-handbook}}. 
We detect the polarized emission in Stokes $Q$, $U$, and $V$ in the continuum, and Stokes $Q$ and $U$ in some channels in the CO(3--2) line. In the Appendix \ref{App:ResInstr}, we discuss the reliability of the polarized detection, taking into account the possible instrumental polarization residuals. In the Appendix \ref{App:SupFig}, we show the detected polarization per velocity channel in the CO(3--2) observations.

%\textcolor{blue}{The maximum liner polarization is about 0.15~\mjyb. Thus, the dust polarization signal is below the detection sensitivity of the channel maps. }

%\textbf{Cuál es la incertidumbre sistemática en la calibración? Referencia a algún trabajo? Discutir posible contaminación, p.ej., fuga de polarización del continuo en la línea?}

%%%%%%%%%%%%%%
\begin{figure*}[ht!]
\includegraphics[width=\textwidth]{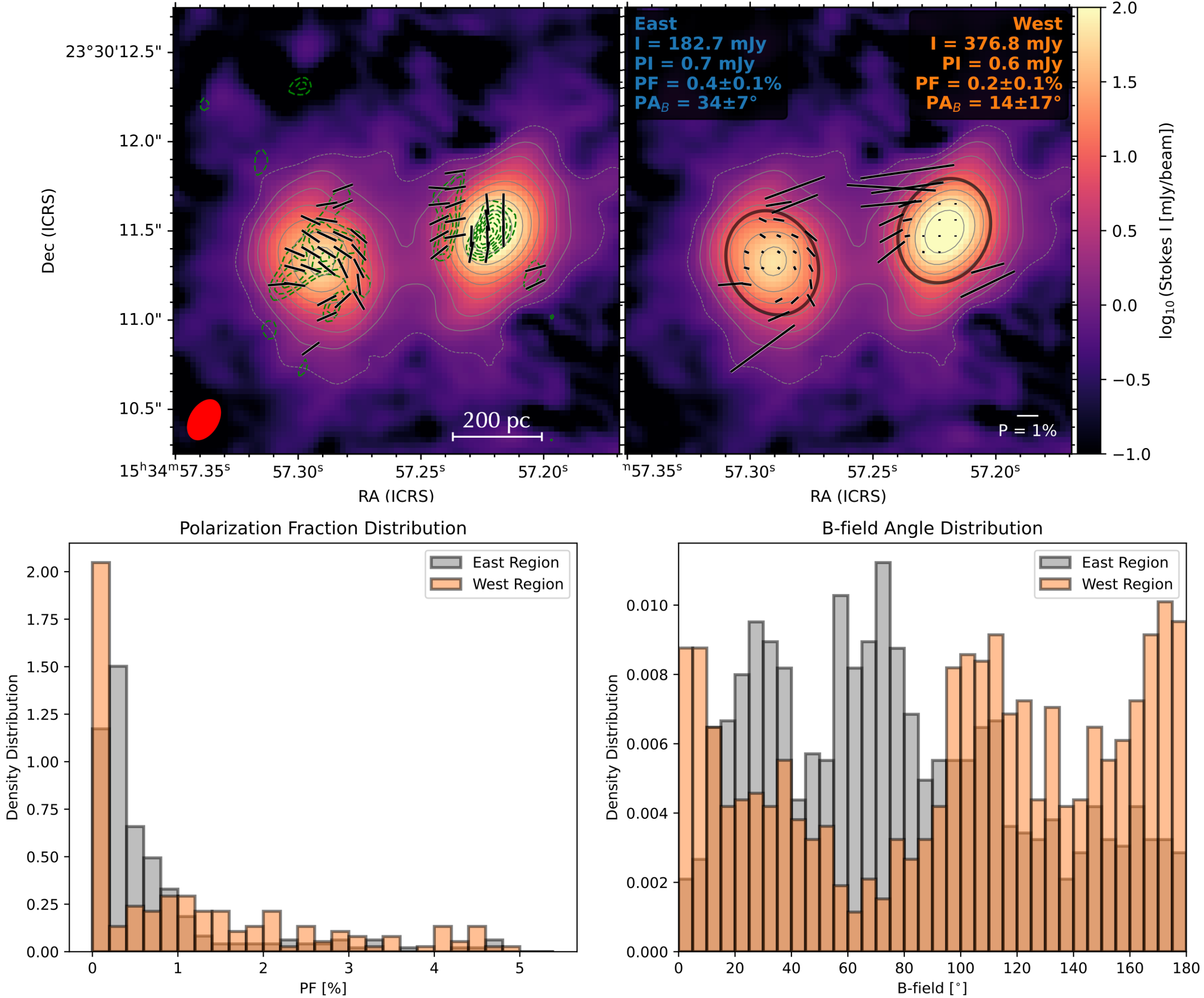}
\caption{The dust continuum polarization and B-field orientation in the central $1\times1$ kpc$^{2}$ of the merger galaxy Arp220 at $345.8$\,GHz ($870\,\mu$m). 
\textit{Top left:} The panel shows the total intensity (I) in colorscale and white contours, and the polarized intensity (PI) in green dashed contours. The contours for I increase as $2^{n}\sigma_{\rm{I}}$, where $n=3,4,5,\dots$, and the contours for PI increase as $n\sigma_{\rm{PI}}$, where $n=4,5,6,\dots$. 
\textit{Top right}: This panel displays the continuum total intensity image (as in the top-left panel), with the length of the black lines representing the polarization fraction (PF). A legend for a $1$\% polarization is provided in the bottom right of the panel. The integrated total intensity, polarized intensity, polarization fraction, and B-field orientation are shown for the East (Arp\,220\,E, cyan) and West (Arp\,220\,W, orange) cores within the apertures (grey ellipses) of $0.5\arcsec\times0.6\arcsec$ ($200\times240$ pc$^{2}$) oriented at $25^{\circ}$ and $-25^{\circ}$, respectively.
In both top panels, the B-field orientation (black lines) is shown at the Nyquist sampling for polarization intensities $\geq 4 \sigma_{PI}$. The synthesized beam (red ellipse) is $0.24\arcsec\times0.16\arcsec$ ($96\times60$ pc$^{2}$), oriented at $-31^{\circ}$.
\textit{Bottom left:} Histogram of the polarization fraction in $0.2$\% bins. 
\textit{Bottom right:} Histogram of the B-field orientation in $5^{\circ}$ bins.
For both histograms, only polarization measurements with $\geq 4\sigma_{\rm{PI}}$ are included. The polarization measurements for the East (grey) and West (orange) cores are shown separately.
 \label{fig:fig2}}
\end{figure*}
%%%%%%%%%%%%%%

%%%%%%%%%%%%%
%%%% TABLE 1 %%%%
%%%%%%%%%%%%%
\begin{deluxetable*}{lcccccc}[ht!]
\tablecaption{Polarimetric measurements. \label{tab:app_table1}}
\tablewidth{0pt}
\tablehead{\colhead{}	& \colhead{Units} & \colhead{East Core} & \colhead{West Core} 
	}
\startdata
\hline
(RA, DEC) & [HH:MM:SS, DD:MM:SS] & (15:34:57.29, 23:30:11.34) & (15:34:57.22, 23:30:11.50) \\
\hline
\multicolumn{4}{c}{Dust Polarized Emission\tablenotemark{a}} \\
\hline
I$_{\rm{cont}}$     & [mJy]         & $182.7\pm18$       &   $376.8\pm38$     \\
PI$_{\rm{cont}}$    & [mJy]         & $0.7\pm0.1$         &   $0.6\pm0.1$       \\
PF$_{\rm{cont}}$    & [\%]         & $0.4\pm0.1$   &   $0.2\pm0.1$ \\
PA$_{B,\rm{cont}}$  &  [$^{\circ}$] & $34\pm7$      &   $14\pm17$   \\
\hline
\multicolumn{4}{c}{ CO(3-2) Polarized Line Emission} \\
\hline
Velocity Range\tablenotemark{b}  &   [km s$^{-1}$]    &   $-$    &   $-462.2$ \\
PF$_{\rm{blue}}$\tablenotemark{c}    & [\%]         & $-$   &   $1.7\pm0.5$ \\
PA$_{E,\rm{blue}}$\tablenotemark{d}  &  [$^{\circ}$] & $-$      &   $-1\pm20$   \\
Velocity Range\tablenotemark{e}  &   [km s$^{-1}$]    &   $-$    &   [$393.1,517.3$] \\
PF$_{\rm{red}}$\tablenotemark{f}    & [\%]         & $-$   &   $[0.7,2.4]$ \\
PA$_{E,\rm{red}}$\tablenotemark{g}  &  [$^{\circ}$] & $-$      &   $81\pm12$   
\enddata
%%\tablecomments{TBD}
\tablenotetext{a}{The median polarization measurements of the dust continuum polarization in the East and West cores within the $0\farcs5\times0\farcs6$ at a PA of $+25^{\circ}$ and $-25^{\circ}$, respectively. The error in the orientation of the B field reflects the scatter of angles within the aperture rather than the uncertainty of measurement.}
\tablenotetext{b}{The velocity range where polarization is detected in the blueshifted outflow (Fig.~\ref{App:fig2}). The blueshifted polarized outflow is unresolved.}
\tablenotetext{c}{The total polarization fraction measured within the blueshifted outflow in the West Core.}
\tablenotetext{d}{The polarization angle (E-vector) within the blueshifted outflow in the West Core. The uncertainty represents the scatter of all individual polarization measurements.}
\tablenotetext{e}{The velocity range where polarization is detected in the redshifted outflow (Fig.~\ref{App:fig2}). The polarized blueshifted outflow is resolved.}
\tablenotetext{f}{The range of polarization fraction within the resolved redshifted outflow in the West Core. The uncertainty of individual polarization measurements is $0.2-0.5$\%, yielding significance levels of $4-6\sigma$ in individual channels.}
\tablenotetext{g}{The median polarization angle (E-vector) of the resolved redshifted outflow in the West Core. The uncertainty represents the scatter of all individual polarization measurements.}
\end{deluxetable*}
%%%%%%%%%%%%%

\section{Results}\label{sec:RES}

We measure different B-field configurations in the cores of Arp\,220 using magnetically aligned dust grains, emission-line polarization, and circular polarization. \revision{For dust polarization, the measured polarization angles were rotated by $90^{\circ}$ to show the plane-on-the-sky B-field orientation.} In Figure~\ref{fig:fig0} we show a summary of our most important results obtained with ALMA and presented and discussed in detail below. Figure \ref{fig:fig2} shows the density distribution of polarization fractions (PF) and B-field orientation for the independent polarization measurements across Arp\,220.

\subsection{Dust polarization of Arp\,220} \label{sec:DustPol}

Arp\,220\,E exhibits a spiral-like B-field structure. In its central region, spanning a diameter of $\sim200$\,pc ($\sim0\farcs5$), the B-field is oriented at $\sim70^{\circ}$, running parallel to the disk of the galaxy \citep{BM2015,Scoville2017} with one arm opening toward the South-West (Figs.~\ref{fig:fig0}, \ref{fig:fig2}). 
Arp\,220\,W has an almost vertical B-field component ($\sim-5^{\circ}$), which is perpendicular to the East-West galaxy disk \citep[Fig. \ref{fig:fig0};][]{BM2015,Scoville2017}. Both the continuum and polarized dust emission are also elongated along this direction, though the central disk of the galaxy remains unresolved in our observations. 
This vertical B-field component and the associated continuum and polarized dust emission structures are parallel to the bipolar molecular outflow in Arp\,220\,W observed by \cite{BM2018}. 
In addition to the galactic structures, we measure a distinct B-field component at $\sim110^{\circ}$ in the region  above the disk of Arp\,220\,E and east of Arp\,220\,W.  We label this the `overlap' region, as its B-field appears to physically connect the two nuclei via a dust bridge. 
\revision{The B-field orientation in the northern region of Arp 220 does not match the spiral structure. This discrepancy may be due to the  combined influence of the northern spiral arm and the overlapping region. Higher spatial resolution observations will be required to resolve both contributions spatially.}
%\revision{Note that the B-fields in the northern region of Arp220\,E require higher spatial resolution observations to spatially resolve the contribution from the northen spiral arm and the overlap region.}
%This $\sim110^{\circ}$ B-field structure is similar to the unresolved polarization measurements of $108\pm15^{\circ}$ and $102.1\pm4.5^{\circ}$ at $850\,\mu$m ($15\arcsec$ aperture; $6$ Kpc) by the JCMT \citep{Seiffert2007} and at $870\,\mu$m ($0.77\arcsec\times0.45\arcsec$; $308\times180$ pc$^{2}$) by the SMA \citep{Clements2025}, respectively, and the $80^{\circ}$ measured at $2.2\,\mu$m ($6\arcsec$; $2.4$ kpc diameter) by \citet{JK1989}.

%\textbf{MPT: sería útil añadir barras de escala (de 100/200 pc) en todas las figuras donde se muestran los núcleos de Arp 220. De hecho, pondría mejor el tamaño de la barra que corresponde al semieje mayor/menor de nuestro haz sintetizado. EN Fig. 1, añadir también el haz sintetizado.}

The polarization fraction across the nuclei ranges from $0.1$ to $5\%$ (Fig. \ref{fig:fig2}; \revision{minimum correspond to the 0.1\% instrumental polarization of ALMA}).
Using an elliptical aperture of $200\times240$\,pc$^{2}$ ($0\farcs5 \times0\farcs6$) centered at the peak of each core and tilted at $+25^{\circ}$ and $-25^{\circ}$ for the East and West nuclei, we compute integrated polarization fractions of $0.4\pm0.1$\% and $0.2\pm0.1$\%, respectively. Table~\ref{tab:app_table1} shows the polarimetric measurements within the elliptical apertures. 
We estimate that both nuclei are nearly optically thin at $345.8$ GHz after obtaining a flux dependence of $\sim\nu^{2.4}$ in the spectral range of $100-691$\,GHz using fluxes within apertures of $0\farcs1-0\farcs5$ from \citet{Sakamoto2008,wilson2014,Scoville2016,Martin2016}.
The core of Arp\,220\,W shows the lowest polarization fraction with $\le0.2$\%. This low polarization may be the effect of the turbulent outflow (Section \ref{sec:COPol}).
For Arp\,220\,E, its core shows a polarization of $\sim0.4$\%, and a maximum polarization of $\sim2$\% at the outskirts of the spiral arms. FIR ($53-214\,\mu$m) polarimetric observations of nearby galaxies using SOFIA/HAWC+ also found an increase in polarization with increasing radius \citep{LR2022}. The observed increase in polarization was attributed to the decrease in column density and the relative increase of ordered B-fields in the outer part of the galaxy. An elongated PI structure is also detected above and below the disk of Arp\,220\,E at a $4-6\sigma$ significance in PI, with polarization fractions of $2-4$\%.
The overlap region has the highest polarization fraction in the range of $3-5$\%. 
%Our resolved polarization structure is compatible with the unresolved polarization measurement of $\sim2.7$\% observed at $345$\,GHz ($870\,\mu$m) by the SMA \citep{Clements2025}.

\subsection{The polarized CO(3-2) molecular outflow of Arp\,220\,W}\label{sec:COPol}

\begin{figure*}[ht!]
\centering
\includegraphics[width=0.9\textwidth]{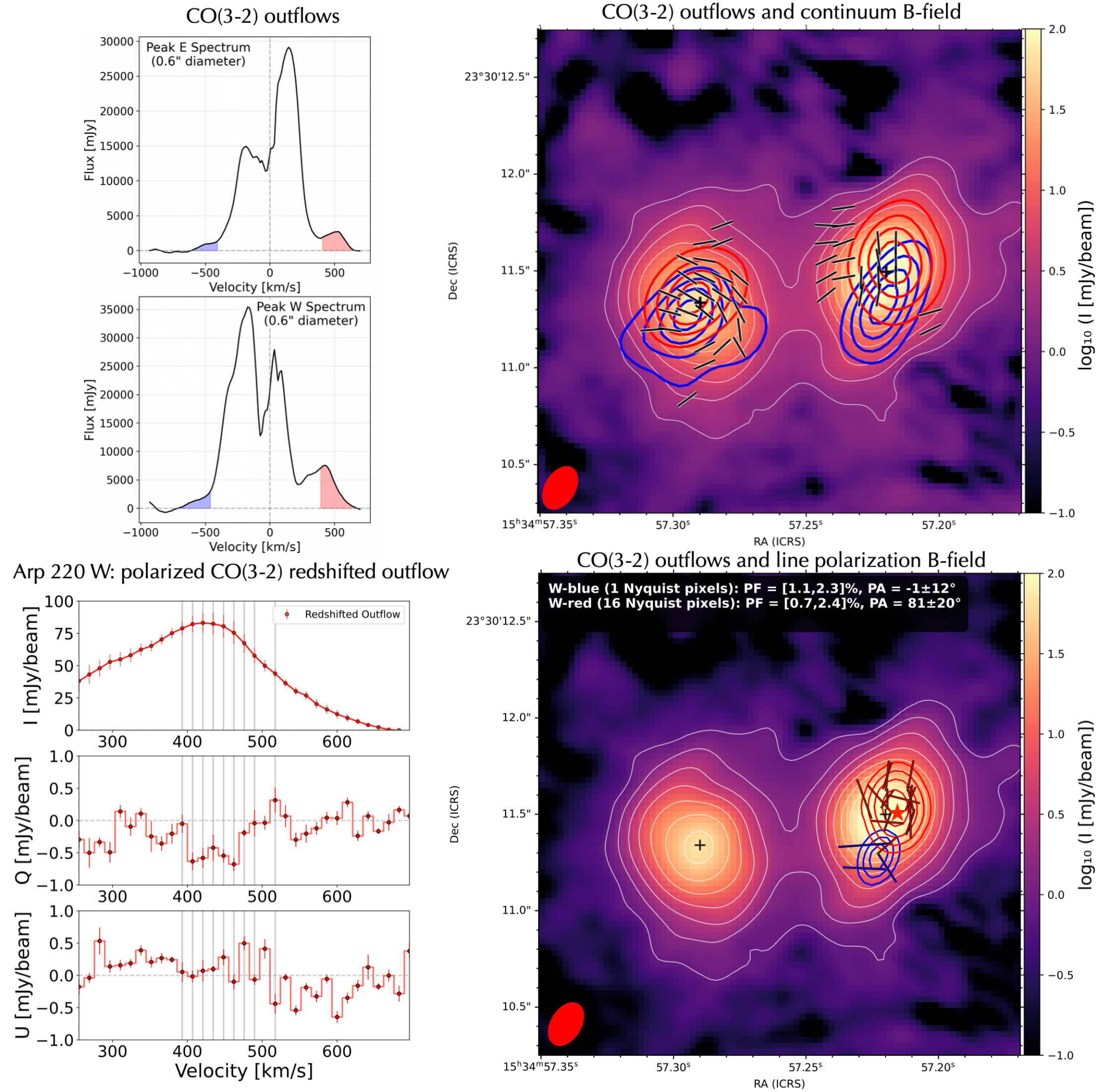}
\caption{The polarized CO(3-2) molecular outflows. 
\textit{Top-left}: the integrated CO(3-2) spectrum of the East (top) and West (bottom) cores, measured within the same aperture as shown in Fig. \ref{fig:fig2}. The plot indicates the rest-frame CO(3-2) line (black dashed line) and the velocity ranges used to define the blueshifted (blue shaded region) and redshifted (red shaded region) outflows.
\textit{Top-right}: Fast molecular redshifted (red contours) and blueshifted (blue contours) outflows are overlaid on the total intensity and B-field orientation, derived from dust polarization data shown in Fig. \ref{fig:fig2}.
\textit{Bottom-left}: The integrated total intensity (top), Stokes Q (middle), and Stokes U (bottom) components for the redshifted outflow. Gray vertical lines denote the velocity channels that have \revision{Nyquist sampled} statistically significant polarization measurements  (Fig.~\ref{App:fig2}).
\textit{Bottom-right}: Line emission polarization measurements are rotated by $90^{\circ}$ to display the B-field orientation of the redshifted and blueshifted outflows over the total dust continuum. \revision{The individual Nyquist sampled polarization measurements per channel that satisfy $PI/\sigma_{PI}\ge4$ are shown (Fig. \ref{App:fig3} shows each channel).} The legend provides key properties for both the blue and redshifted outflows, including the range of polarization, the median position angle of polarization, the scatter of the position angle within the outflow, and the number of independent Nyquist-sampled polarization measurements.
 \label{fig:fig3}}
\end{figure*}

The most remarkable result is the detection of emission line polarization along the CO(3-2) molecular outflow of Arp\,220\,W (Figs. \ref{fig:fig0} and \ref{fig:fig3}). We show the integrated intensity of the CO(3-2) molecular outflows for each core, using the velocity ranges as in \citet{Wheeler2020}, which reveals the known fast ($\sim500$ km s$^{-1}$) molecular outflows in both cores. For each outflow velocity channel, we identify the polarization in those pixels that satisfy the condition $PI/\sigma_{\rm{PI}} \ge 4$. The individual velocity channels with polarization measurements satisfying these conditions are shown in Appendix \ref{App:SupFig} (Fig. \ref{App:fig2}). Table~\ref{tab:app_table1} shows the velocity channels with detected polarized flux and the polarization measurements for each outflow. We measure one independent beam at a velocity of $-462$ km s$^{-1}$ in the blueshifted outflow, and 16 independent beams within the velocity range of $[393.1,517.3]$ km s$^{-1}$ in the redshifted outflow. For simplicity, Fig. \ref{fig:fig3} shows the polarization measurements of each velocity channel of the blueshifted and redshifted outflows. These polarization measurements are overlaid on the integrated  flux of the CO(3-2) outflow associated with the velocity channels with detected polarized flux. 

The CO(3-2) molecular outflow of Arp\,220\,W is polarized at levels of $[1.1-2.3]$\% and $[0.7-2.4]$\% in the blue and redshifted outflows, respectively. Each of these measurements has a significance of $\ge 4\sigma_{PI}$ in its corresponding velocity channel (Fig. \ref{App:fig2}). The median polarization fractions are $1.7\pm0.5$\% and $1.2\pm0.3$\% in the blue and redshifted outflows, respectively. The median position angles of polarizations are $-1^{\circ}$ and $81^{\circ}$, with angular variations of $12^{\circ}$ and $20^{\circ}$ across the blue and redshifted outflows, respectively.

%%%%%%%%%%%%%%
%%%%%%%%%%%%%%

\subsection{Circular polarization in the core of Arp\,220\,E}\label{sec:StokesV}

We report the likely detection of circular polarization (Stokes V) of $0.3\pm0.1$\%, in Arp\,220\,E within a $0\farcs5$ ($200$\,pc) aperture. We analyze the residual polarization as a function of the paralactic angle during the 4h of observations and measure a persistent Stokes V emission in the Arp\,220\,E, which suggests that the Stokes V may be real. However, the Stokes V in the Arp\,220,W is consistent with instrumental polarization. Appendix \ref{App:ResInstr} describes the details of this analysis. Note that  circularly polarized dust emission has been reported in the near-IR \citep{Chrysostomou2007, Kwon2014, Kwon2016}. As Arp\,220\,E is linearly polarized, the circular polarization can be interpreted as the conversion from linear to circular polarization by a foreground of dust grains aligned by a B-field. Given the uncertainty in the reliability of the detection, however, we do not consider this option further.

%\textbf{MPT:  
%\begin{itemize}
%    \item Cuáles son las implicaciones físicas de un campo magnético paralelo vs. perpendicular al flujo? 
%    \item Se podría/debería comparar con flujos galácticos, e.g. protoestelares? 
%    \item Estimar las energías magnética y cinética del flujo y compararlas.
%    \item Discutir los efectos de depolarización, efecto de "beam-smearing" en los resultados?
%\end{itemize} }

\section{Summary and Discussion}
\label{sec:DIS}

We summarize the main findings of our paper as follows.

\paragraph{(a) Dust polarized outflows in Arp\,220\,W.} Arp~220~W shows a predominantly vertical B-field ($\mathrm{PA} \approx -5^{\circ}$; Figs.~\ref{fig:fig0}, \ref{fig:fig2}) that is aligned with the bipolar molecular outflow \citep{BM2018}. The polarization fraction is $P = 0.2 \pm 0.1\%$ across the outflows. The observed low polarization fraction likely results from depolarization due to the highly turbulent outflow ($\sim220$~km~s$^{-1}$; Fig.~\ref{App:fig1}) and tangled B-fields along the line of sight. Nevertheless, the polarized signal is statistically significant, indicating an ordered B-field component. Higher angular resolution could resolve the turbulent coherence length and reveal a more complex structure. Similar alignments between B-fields and outflows have been observed in nearby starburst galaxies, where the cold, dense medium shows magnetic fields parallel to the galactic outflows and extending tens of kiloparsecs from the disk \citep{LR2021,Pattle2021,LR2022,LR2023,Borlaff2023}. While most galaxies exhibit B-fields parallel to their disks \citep{Beck2013,Borlaff2023}, the combined effects of flux freezing, vertical shear, and pressure minimization can drag the B-field into alignment with the outflow \citep{LR2021}. Our results for Arp~220 follow this general pattern.

\paragraph{(b) Polarized CO(3--2) molecular line outflows in Arp\,220\,W.} We report the first detection of CO(3--2) line polarization in the outflow of an external galaxy. The polarization fractions range between $1-2$\% with a statistical significance of $4-6\sigma$ in independent velocity channels. The blueshifted and redshifted outflows exhibit median polarization angles of $\sim-1^{\circ}$ and $\sim81^{\circ}$, respectively. However, line emission polarization angle has a $90^{\circ}$ ambiguity, and a detailed radiative transfer analysis using the physical conditions of the outflow is required to estimate the exact orientation of the B-field in the molecular outflow \citep[e.g.,][]{Lankhaar2020,Lankhaar2022,Lankhaar2023}. The dust polarization, from magnetically aligned dust grains, provides an alternative approach to estimate the B-field orientation of the CO(3-2) molecular outflow. We estimate an angular offset between the polarization angle from thermal and molecular emission of $|-5^{\circ} - 81^{\circ}| = 89^{\circ}$. Given the angular variations in both molecular line emission ($\sim20^{\circ}$) and dust polarization ($\sim17^{\circ}$), the measured molecular line emission polarization is most likely to be oriented $90^{\circ}$ from the B-field orientation across the outflow. The main assumption is that both tracers are sensitive to the dominant B-field component in the outflow, i.e., parallel to the outflow as directly measured by dust polarization. Thus, a $90^{\circ}$ rotation has been applied to the CO(3--2) polarization angles displayed in Figs.~\ref{fig:fig0} and~\ref{fig:fig3}. This configuration suggests that the $\sigma M=\pm1$ (polarization perpendicular to the B-field) transitions are overpopulated in the outflow relative to the $\pi M = 0$ (polarization parallel to the B-field) transitions. We conclude that both dust polarization and CO(3-2) molecular line polarization are most likely to show a B-field parallel to the galactic outflow.

%These angles are offset by about $90^{\circ}$ relative to the dust polarization, consistent with the Goldreich--Kylafis effect, in which the $\sigma_M = \pm 1$ transitions (polarization vectors perpendicular to the magnetic field) dominate. This alignment suggests that the magnetic field in the molecular gas is parallel to the large-scale galactic outflow.

%The comparison between thermal dust and molecular-line polarization confirms an angular offset of $|-5^{\circ} - 81^{\circ}| = 89^{\circ}$. Given the observed angular variations in both tracers ($>22^{\circ}$ for CO and $\sim10^{\circ}$ for dust), the measured CO(3--2) polarization is most likely oriented $90^{\circ}$ from the magnetic-field direction across the outflow. Assuming both tracers probe the same dominant field component, we infer that the B-field traced by CO emission is parallel to the outflow axis. A $90^{\circ}$ rotation has therefore been applied to the CO polarization angles  displayed in Figs.~\ref{fig:fig0} and~\ref{fig:fig3}.

%%%%%%%%%%%%%
%%%% TABLE 2 %%%%
%%%%%%%%%%%%%
\begin{deluxetable*}{lllcccc}[ht!]
\tablecaption{Magnetic field strengths in Arp\,220\,W \label{tab:table2}}
\tablewidth{0pt}
\tablehead{\colhead{Symbol}	& \colhead{Description} & \colhead{Equation} & \colhead{Redshifted} & \colhead{Blueshifted} 
	}
\startdata
\hline
\multicolumn{5}{c}{This work} \\
\hline
B$_{\rm{eq}}$   &   Turbulent kinetic and B-field energies equipartition & $B_{\rm{eq}} = \sqrt{4\pi\rho_{\rm{CO}}}\sigma_{\rm{CO}}$ & $6.4\pm1.6$ mG    &   $2.3\pm0.6$ mG   \\
B$_{DCF}$       &   Classical DCF method &  $B_{\rm{DCF}} = \sqrt{4\pi\rho_{\rm{CO}}}\frac{\sigma_{\rm{CO}}}{\sigma_{\rm{\theta}}}$ & $18.2\pm6.8$ mG &   $4.0\pm1.7$ mG \\
B$^{\prime}_{DCF}$ & Modified DCF in outflows &   $B^{\prime}_{\rm{DCF}} = B_{\rm{DCF}}|1- \sigma_{\theta}\frac{U_{0}}{\sigma_{v,\rm{CO}}}|$ & $9.5\pm5.6$ mG & $1.1\pm1.0$ mG \\
\hline
\multicolumn{5}{c}{From literature\tablenotemark{a}} \\
\hline
B$_{\rm{ISM}}$  &   Volume average in ISM   &  $-$    &  \multicolumn{2}{c}{$0.9$ mG} \\
B$_{\rm{OH}}$        &   LOS B-field in OH masing clouds (Arp\,220\,W) &   $-$   & \multicolumn{2}{c}{$1-8$ mG}\\
B$_{\rm{min}}$      &   Cosmic ray density and B-field equipartition &   $-$  & \multicolumn{2}{c}{$0.24$ mG} \\
B$_{\rm{SNR}}$       &   B-field in ISM estimated from synchrotron in SNR    &   $-$   & \multicolumn{2}{c}{$2.7$ mG} \\
B$_{\rm{SFR}}$ & B-field enhanced by SFR (Arp\,220\,W) & $B\propto SFR^{0.34}$ & \multicolumn{2}{c}{$0.66$ mG (core); $0.22$ mG (outflow)}
\enddata
\tablenotetext{a}{
B$_{\rm{eq}}$, B$_{\rm{OH}}$, B$_{\rm{Bmin}}$, and B$_{\rm{SNR}}$ are taken from \citet{McBride2014}.
B$_{\rm{SFR}}$ uses the scaling factor from M\,82 estimated by \citet{LR2023}. See Section \ref{sec:DIS}-c for details.}
\end{deluxetable*}
%%%%%%%%%%%%%

\paragraph{(c) Magnetic-field strengths of the outflow} 
We provide estimations of the B-field strengths in the outflow of Arp\,220\,W assuming several approaches (Table~\ref{tab:table2}). 

We estimate the B-field in equipartition with the turbulent kinetic energy, such as $B_{\rm{eq}} = \sqrt{4\pi\rho_{\rm{CO}}}\sigma_{\rm{CO}}^2$, where $\rho_{\rm{v,CO}}$ is the density and $\sigma_{v,\rm{CO}}$ is the three-dimensional dispersion velocity of the polarized outflow defined as $\sqrt{3}\sigma_{\rm{CO}}$. As the molecular phase typically dominates the mass outflow rate budget,  which exceeds the ionized gas phase by 1-2 orders of magnitude \citep[e.g.,][]{GB2015,Fiore2017,Cresci2023}, and since we trace similar B-fields in both dust and molecular gas, we use the physical properties of the molecular gas in the outflows. We measure a velocity dispersion of $\sim220$ km s$^{-1}$ in the polarized molecular outflows and take the CO(3-2) molecular masses of $10^{6}$ M$_{\odot}$ and $10^{7}$ M$_{\odot}$ within a volume of $90\times50\times50$ and $121\times50\times50$  pc$^{3}$ (length$\times$width$\times$depth; assuming an axysymmetric outflow) in the blue and redshifted outflows, respectively, from \citet{Wheeler2020}. \revision{Table \ref{tab:app_table1} shows these values and their uncertainties. We compute the median and standard deviation using bootstrapping with 1,000 random iterations for the blue and redshifted outflows.} Thus, we estimate equipartition B-field strengths of $2.3\pm0.6$ mG and $6.4\pm1.6$ mG in the blue and redshifted outflows. 
\revision{The mass is the dominant source of the overall uncertainty, introducing an uncertainty of a factor of two \citep{Wheeler2020}.}
%\revision{The highest contributor to the uncertainty is the mass, which has a factor two uncertainty \citep{Wheeler2020}.}

To estimate the B-fields produced by magnetically aligned dust grains, we follow the approach developed by \citet{LR2021} to estimate the B-fields along the galactic outflow of the starburst galaxy M\,82. \revision{Table \ref{tab:app_table1} shows the quantities used to estimate the B-field strengths. The angular dispersion, $\sigma_{\theta}$, was estimated as the standard deviation of the polarization angles of the dust and molecular line polarization measurements, which intrinsically assume that both the dust and molecular gas are cospatially embedded in the same B-field of the outflow. Specifically, the redshifted outflow has 4 and 16 Nyquist measurements for dust and molecular line polarization, respectively, and the blueshifted outflow has 2 and 4 Nyquist measurements for dust and molecular line polarization, respectively. Higher angular resolution observations are required to spatially resolve the outflows and enable the use of more sophisticated approaches \citep[e.g., structure function;][]{Hildebrand2009}, so our measurements should be considered upper limits and order-of-magnitude estimates}. We estimate B-field strengths of \revision{$4.0\pm1.7$\,mG and $18.2\pm6.8$\,mG} for the blue and redshifted outflows, respectively, using the Classical Davis-Chandrasekhar-Fermi (DCF) method \citep{Davis1951,CF1953}. \revision{Both the mass and the angular dispersion equally contribute to the uncertainty}. However, the Classical DCF method cannot be applied to outflows due to the effects of shearing and compression. These B-field strengths should be considered as upper-limits.  \citet{LR2021}, further supported by \cite{Guerra2023} using MHD simulations and solving ideal-MHD equations from first principles, considering the effects of a large-scale, shear flow on the propagation of a fast magnetosonic wave, derived a modified DCF applicable to outflows (Table~\ref{tab:table2}). We use the outflow velocities of \revision{$U_{0} = 460$ and $500$ km s$^{-1}$, and estimate B-field strengths of $1.1\pm1.0$ mG and $9.5\pm5.6$ mG} in the blue and redshifted outflows, respectively. 
\revision{Despite the large uncertainties, the difference in their B-field strengths suggests that the discrepancy is probably real.}
%\revision{The highest contributor to the uncertainty is the angular dispersion followed by the mass and the velocity dispersion of the molecular gas.} 
%\revision{Despite the large uncertainties in the estimation of the B-field strengths, the difference is significant at a $3\sigma$ level, which indicates that a physical mechanism produces the difference.} 
The stronger B-field in the redshifted outflow arises from its larger mass (by a factor of ten) and smaller angular dispersion (by a factor of two).

Our estimated magnetic-field strengths are comparable with previous measurements using independent techniques across the multi-phase medium of Arp\,220 (Table~\ref{tab:table2}). 
Zeeman splitting measurements provide the most direct method to estimate B-fields, which were found to be in the range of $1-8$ mG in the core of Arp\,220\,W \citep{McBride2015}. This B-field is close to the equipartition with both turbulent kinetic energy and the gravitational energy of the masing clouds. 
The B-field estimated from synchrotron emission in supernova remnants (SNR) of Arp\,220 was measured to be $2.7$ mG. This B-field is generated by post-shock B-field amplification, which results from the shock compression of the ambient ISM \citep{Thompson2009}. Indeed, these B-field strengths are larger than the volume-averaged B-field in the ISM of 0.9~mG, and 0.24~mG assuming the cosmic rays and B-field equipartition \citep{McBride2014}. 

The B-field strength is correlated with the star formation rate (SFR) as $B\propto\,SFR^{0.34}$ \citep{VanEck2015,Beck2019}. Using the FIR B-field strength of M\,82, $305\pm5\,\mu$G \citep{LR2021}, the scaling factor is $B_{0} = 128\,\mu$G M$_{\odot}^{-0.34}$ yr $^{-0.34}$ \citep{LR2023}. The surface SFR in Arp\,220\,W is $10^{4.1}$ M$_{\odot}$ yr$^{-1}$ Kpc$^{-2}$. Using an area of the core of $200\times50$ pc$^{2}$ (Fig. \ref{fig:fig0}), this yields an SFR of  $125$ M$_{\odot}$ yr$^{-1}$ at the core of Arp\,220\,W. The SFR within the outflow of Arp\,220\ is 5~M$_{\odot}$ yr$^{-1}$ \citep{Perna2020}. Using the correlation between B-field and SFR, we estimate B-field strengths of $0.66$ mG and $0.22$ mG for the core and the outflow of Arp\,220\,W, respectively. Although the B-fields derived from the SFR reach hundreds of $\mu$G (which can be the driving mechanism in the ISM), they do not fully explain the measured mG in the masing clouds and outflows. The discrepancy between these predictions and the observed mG B-fields indicates substantial amplification due to compression within the nuclear clouds and post-shocks from SNRs. We conclude that our estimated B-field in the dusty and molecular outflow is likely due to an amplified B-field driven by compression in the nuclear clouds and post-shock compression by SNRs. This amplified B-field is then sustained by the turbulent kinetic energy in the outflow and dragged away into the CGM.

\paragraph{(d) Spiral B-fields in Arp\,220\,E.} The B-field follows a spiral-like structure with a mean position angle of $\sim70^{\circ}$ with the spiral arm opening towards the south-western region. Spiral B-fields are common in spiral galaxies \citep{Beck2013,LR2022,Borlaff2023} and are attributed to differential rotation driven by the large-scale dynamo mechanism \citep{Subramanian1998,BS2005}. As in the case of the B-fields in the Antennae galaxy \citep{LR2023b}, also a merger system, our observations show that the relic spiral arm still has an ordered B-field that can persist during the current merger phase. 

\paragraph{(e) Polarized dusty bridge.} Between both nuclei, in the `overlap' region, we detect the highest polarization levels, $P = 3-5\%$, with orientations $\sim110^{\circ}$. This configuration is consistent with the unresolved polarization measurements previously reported with the James Clerk Maxwell Telescope  \citep[JCMT;][]{Seiffert2007} and the Submillimeter Array  \citep[SMA;][]{Clements2025}, which may have been sensitive to the dust ridge connecting the two nuclei.

We emphasize that Arp~220 serves as the nearest prototype of an ultra-luminous infrared galaxy and provides a benchmark for understanding feedback processes in high-redshift systems. The $\sim$mG B-fields detected here are comparable to or stronger than those inferred in distant dusty star-forming galaxies. B-field strengths $\lesssim500\,\mu$G have been reported in a fast ($\sim300$~km~s$^{-1}$) rotating disk at $z=2.6$ (9io9; \citealt{geach2023}) and in the overlap region of a merger at $z=5.6$ (SPT0346--52; \citealt{Chen2024}), both with SFRs near $10^3$~M$_\odot$~yr$^{-1}$. The Arp~220 results suggest that such strong B-fields may be common in extreme starbursts, amplified by turbulence, shear, and feedback. Furthermore, our detection of CO(3--2) molecular-line polarization demonstrates that this diagnostic can be applied to distant galaxies. Future ALMA observations targeting similar transitions in high-$z$ systems could reveal widespread $\sim100$--$1000$~$\mu$G B-fields in molecular outflows, confirming that magnetic fields play a fundamental role in regulating star formation and feedback across cosmic time.

%% Please use the acknowledgment and contribution environments. This will 
%% be anonomyized when the "anonymous" style option is used. 

\begin{acknowledgments}
This paper makes use of the following ALMA data: ADS/JAO.ALMA\#2023.1.00044.S. ALMA is a partnership of ESO (representing its member states), NSF (USA) and NINS (Japan), together with NRC (Canada), NSTC and ASIAA (Taiwan), and KASI (Republic of Korea), in cooperation with the Republic of Chile. The Joint ALMA Observatory is operated by ESO, AUI/NRAO and NAOJ.
The National Radio Astronomy Observatory is a facility of the National Science Foundation operated under cooperative agreement by Associated Universities, Inc.

E.L.-R. thanks support by the NASA Astrophysics Decadal Survey Precursor Science (ADSPS) Program (NNH22ZDA001N-ADSPS) with ID 22-ADSPS22-0009 and agreement number 80NSSC23K1585. 
G.B., J.M.G., and J.M.T. acknowledge support from the PID2023-146675NB grant funded by MCIN/AEI/10.13039/501100011033. M.M. acknowledges support from the Spanish Ministry of Science and Innovation through the project PID2021-124243NBC22. J.M.G., M.M., and J.M.T. are also supported by the programme Unidad de Excelencia Mar\'{\i}a de Maeztu CEX2020-001058-M.
A.A. and M.P.-T. acknowledge financial support from the Severo Ochoa grant CEX2021-001131-S and from the Spanish grant PID2023-147883NB-C21, funded by MCIU/AEI/ 10.13039/501100011033, as well as support through ERDF/EU.

\end{acknowledgments}

%\begin{contribution}
%%This section gives authors the space to recognize author contributions. The text inside this environment is NOT counted towards the total word quanta. At a minimum, manuscripts are expected to include this text:

%E.L.-R. was responsible for the data analysis of the ALMA polarization data in continuum and CO, writing, and submitting the manuscript.

%% But authors are expected to provide more specific details, e.g. 
%%
%%SC was responsible for writing and submitting the manuscript.
%%WWM came up with the initial research concept and edited the manuscript.
%%OTS obtained the funding and edited the manuscript.
%%EBF provided the formal analysis and validation. He also edited the manuscript.
%%GEH Supervised the undergraduates, wrote the software and administers the project github and Zenodo repositories.
%%
%% Authors can use the Contributor Role Taxonomy (CRediT) at
%% for ideas on how write a good statement tailored to their needs.

%\end{contribution}

\facilities{ALMA}

\software{astropy \citep{astropy2013,astropy2018,astropy2022},  
          }


\bibliography{references}{}
\bibliographystyle{aasjournalv7}

\vfil\eject

\appendix

\restartappendixnumbering

\section{On the residual instrumental polarization}\label{App:ResInstr}

In this section, we analyze the possible effect of residual instrumental polarization. Chapter 8 of the ALMA technical book\footnote{ALMA technical book: \url{https://almascience.nrao.edu/proposing/technical-handbook}} indicates that the main source of on-axis instrumental polarization is caused by D-term leakages. These leakages are more significant in the cross-products, $XY$ and $YX$, as they are dominated by the D-term values multiplied by the parallel hands, $XX$ and $YY$.
After correcting for leakages, the expected accuracy in polarization calibration should be $0.1\%$ for linear polarization and $1.8\%$ for circular polarization.
The leakages are estimated on-axis. However, there is an additional off-axis residual instrumental polarization that increases with distance from the phase center \citep{Hull2020}. Considering the distance of the cores to the phase center (\revision{approximately $0\farcs5$ for both 
cores}), 
the expected off-axis instrumental polarization in Band 7 is insignificant ($<<0.1\%$) for linear polarization, but is approximately $0.10-0.15\%$ for Stokes $V$ due to beam squint \citep[Fig.~A.1 and 4 from][]{Hull2020}. In both cases, the instrumental polarization is sensitive to the parallactic angle, whereas the real signal from Stokes $Q$, $U$, and $V$ is not.

\begin{figure*}[ht!]
\includegraphics[width=0.39\textwidth]{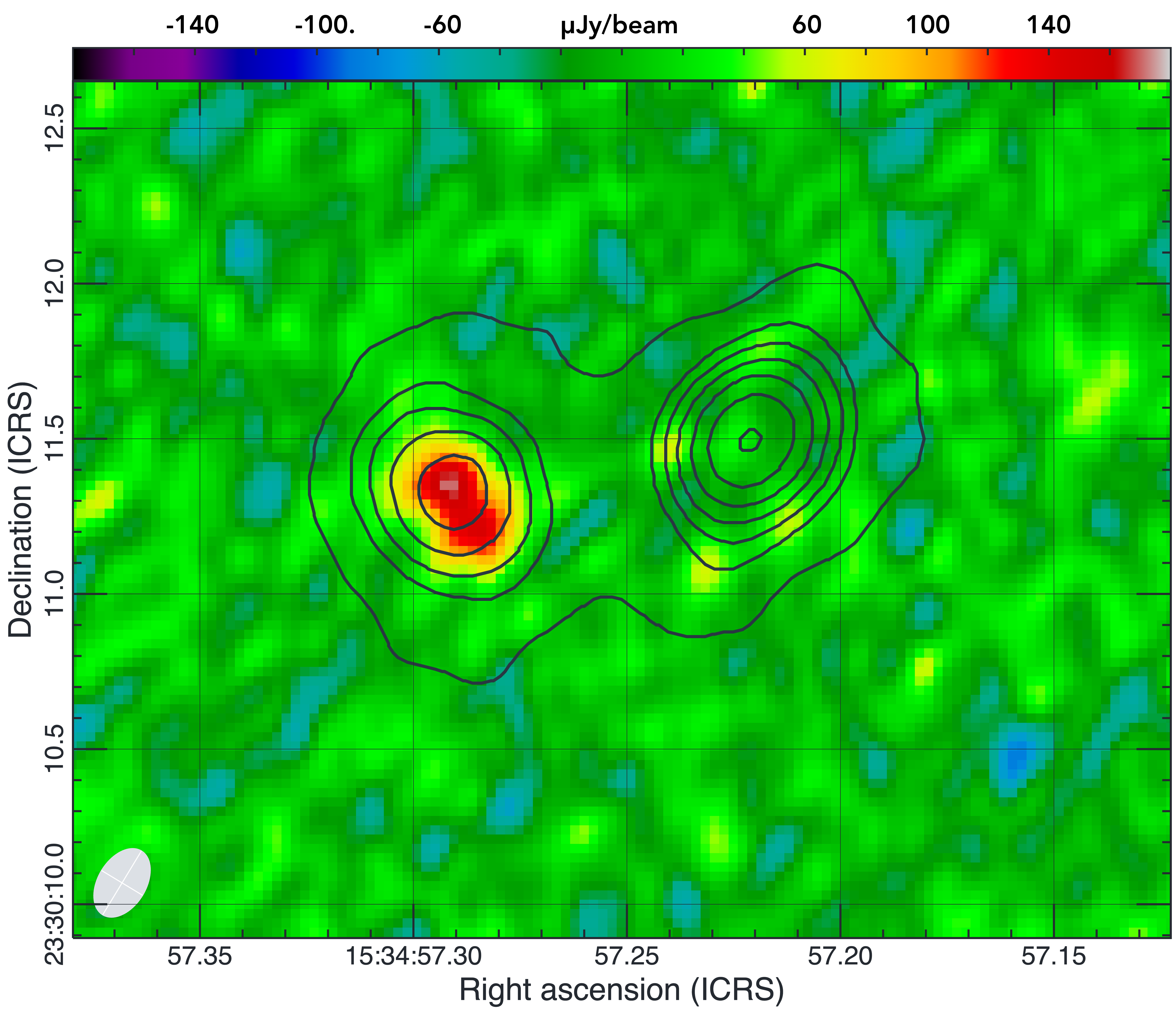}
\includegraphics[width=0.61\textwidth]{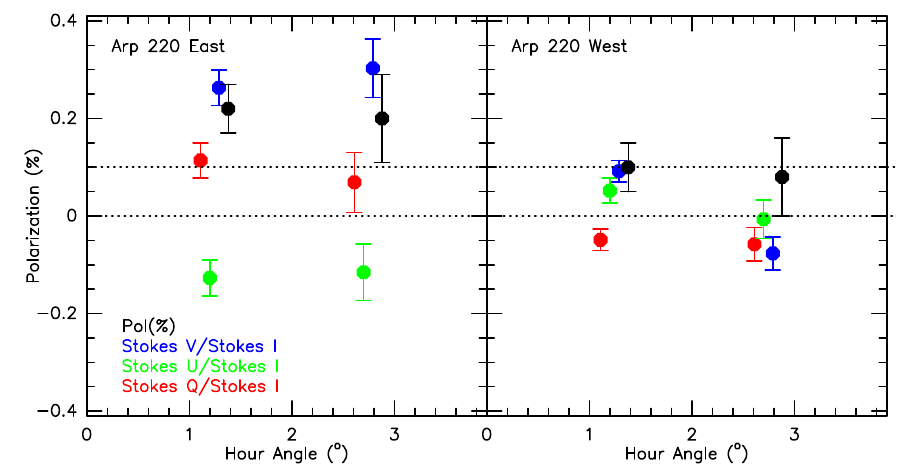}
\caption{{\it Left panel:} A colour image of Stokes $V$ overlaid with Stokes $I$ contours. Contours are at levels of 1, 5, 20, 40, 80 and 160 mJy~beam$^{-1}$.
{\it Center and Right panels:}
The fraction (in percentage) of the Stokes $Q$ (green), $U$ (red), $V$ (blue) flux with respect to the Stokes $I$  flux for Arp\,220\,E and Arp\,220\,W. The data are shown for two different HA intervals: 0.0h to 2.0h and  2.0h to 3.4h. The linear polarization fraction (black) is also displayed. The two horizontal dashed lines mark the 0.0 and 0.1\% polarization levels.
 \label{App:fig3}}
\end{figure*}

The linear dust polarization levels of the nuclei are low, below $1\%$, particularly in the West core (Table~\ref{tab:app_table1}), but detected at a statistically significant level above the expected accuracy in linear polarization. Additionally, we measure a significant Stokes $V$ signal, with a peak intensity of $0.17\pm0.02$ $\text{mJy} \text{beam}^{-1}$, or $0.28\%$, in the East core, while the West core does not show significant emission at the $3\text{-}\sigma$ level ($\lesssim0.01\%$, Fig.~\ref{App:fig3}). One would expect the off-axis instrumental polarization and residual leakages to affect the West core more, as it is approximately two times brighter than the East core. The absence of notable circular polarization in the West core, which falls below the expected level of off-axis instrumental polarization, could be attributed to the parallactic angle coverage during observations (from $\simeq 125\arcdeg$ to $180\arcdeg$). This extensive range may cause the off-axis instrumental polarization to become significantly smeared or canceled out. To investigate the possible contributions of the residual instrumental polarization, we split the visibilities into two different Hour Angle (HA), or parallactic angle, ranges: from $0.0\text{h}$ to $2.0\text{h}$ ($140\arcdeg-180\arcdeg$) and from $2.0\text{h}$ to $3.4\text{h}$ ($125\arcdeg-140\arcdeg$). We used the same aperture as in Section~\ref{sec:DustPol} to measure the flux densities of the Stokes parameters. Figure~\ref{App:fig3} shows that the Stokes $Q$, $U$, and $V$ values for the two HA ranges are, within uncertainties, the same for the East core. This consistency is also true for Stokes $Q$ in the West core, but there is a marginal discrepancy for Stokes $U$. However, the Stokes $U$ signal is weaker than Stokes $Q$ (in the $2.0\text{h}-3.4\text{h}$ HA range, Stokes $U$ is only marginally detected), and thus the linear polarization is dominated by Stokes $Q$. The linear polarization fraction in both cores is, within uncertainties, consistent. Stokes $V$ in the West core is detected in both HA ranges at a level of $0.08\%$ with respect to Stokes $I$. This value is about three times weaker than in the East core and is consistent with the expected value for off-axis beam-squint, suggesting this signal is likely instrumental. Crucially, the Stokes $V$ emission between the two HA ranges has opposite signs, implying that they will tend to cancel out in the combined final image.

In summary, the linear polarization detected in the East core is clearly real and does not show signs of residual instrumental effects. In the case of the West core, the linear polarization fraction is at the level of the ALMA accuracy, but the consistency between the two HA ranges suggests that it appears to be real, although we cannot definitively rule out a slight influence from instrumental polarization. Regarding the circular polarization, the measured values are much lower than the ALMA accuracy ($1.8\%$). However, the persistent Stokes V emission in the East core (in contrast to the West core emission) suggests that it might be real. 
%Detection of circularly polarized dust emission has been previously reported in the near-IR \citep[e.g.,][]{Chrysostomou2007, Kwon2014, Kwon2016}. This phenomenon can be caused by various mechanisms, such as the conversion from linear to circular polarization by a foreground of grains aligned by a magnetic field. Given the uncertainty in the reliability of the detection, however, we do not consider this option further.
 
\vfil\eject

\restartappendixnumbering

\section{B-field strength parameters}\label{App:Bfield}

Table \ref{tab:app_table1} shows the parameters and their uncertainties used to estimate the B-field strengths shown in Table \ref{tab:table2} and described in Section \ref{sec:DIS}-c.

%%%%%%%%%%%%%
%%%% TABLE 3 %%%%
%%%%%%%%%%%%%
\begin{deluxetable*}{llccccc}[ht!]
\tablecaption{Parameters to estimate the magnetic field strengths in Arp\,220\,W \label{tab:app_table1}}
\tablewidth{0pt}
\tablehead{\colhead{Symbol}	& \colhead{Description} & \colhead{Redshifted} & \colhead{Blueshifted} 
	}
\startdata
\hline
$\sigma_{\rm{CO}}$  & Velocity dispersion CO(3-2)           &   $220\pm20$ km s$^{-1}$  &   $220\pm20$ km s$^{-1}$ \\
M$_{\rm{CO}}$\tablenotemark{a}       & Molecular mass in outflow             &   $10^{7}$\,M$_{\odot}$   &   $10^{6}$\,M$_{\odot}$ \\
V\tablenotemark{b}                   & Volume of outflow (height $\times$ width $\times$ depth)  &   $121\times50\times50$ pc$^{3}$ & $90\times50\times50$ pc$^{3}$ \\
U$_{o}$             & Velocity molecular outflow CO(3-2)    &   $500\pm50$ km s$^{-1}$  &   $460\pm50$ km s$^{-1}$ \\
$\sigma_{\theta}$   & Angular dispersion                    &   $12\pm5^{\circ}$        &   $20\pm10^{\circ}$ \\
\enddata
\tablenotetext{a}{Molecular mass in the outflows from \citet{Wheeler2020}. We assume a factor 2 in uncertainty.}
\tablenotetext{b}{We assume an axysymmetric outflow.}
\end{deluxetable*}
%%%%%%%%%%%%%

\section{Supplementary figures}\label{App:SupFig}

\begin{figure*}[ht!]
\includegraphics[width=\textwidth]{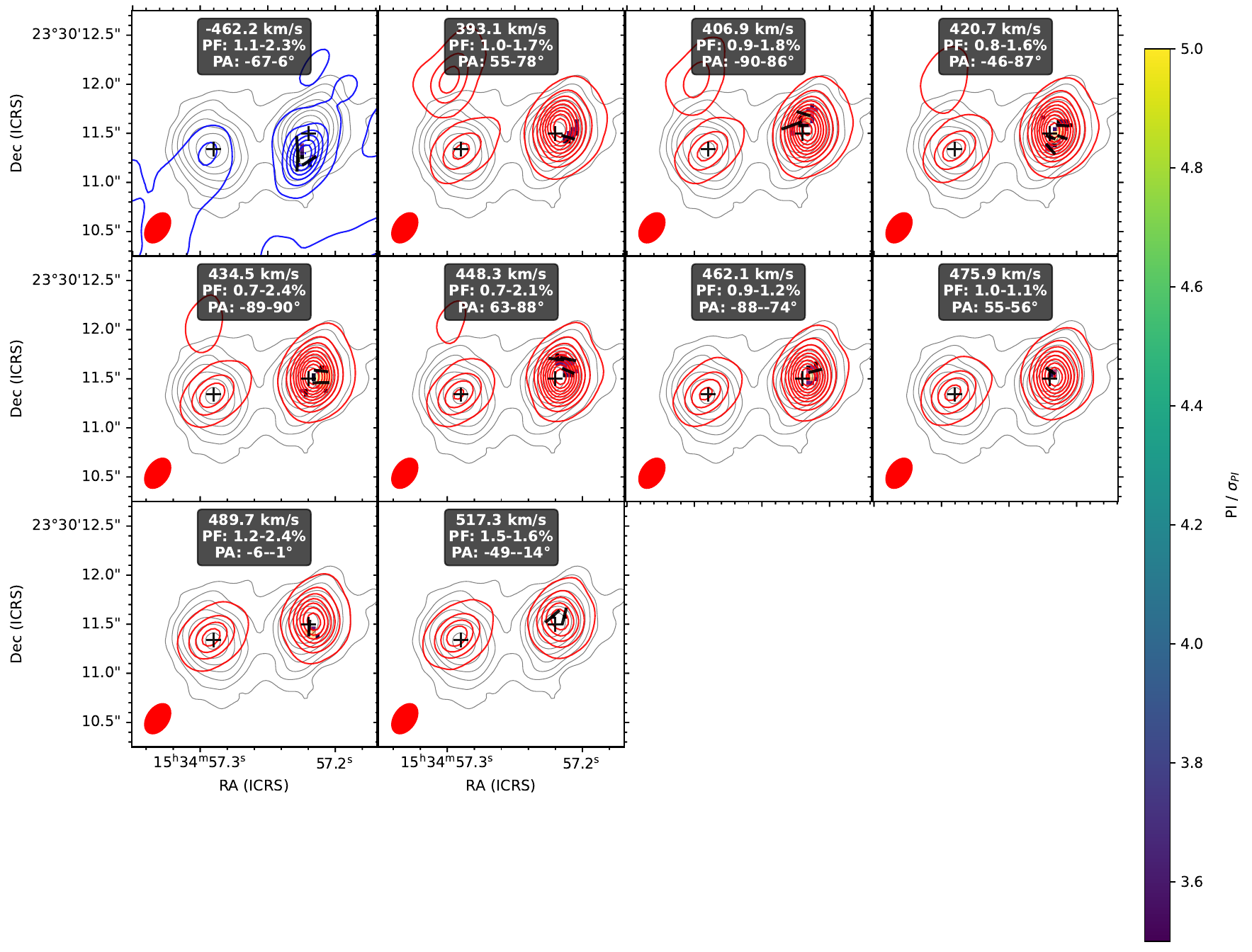}
\caption{Detected CO(3--2) polarization maps per channel in the outflows within the central $1\times1$\,kpc$^{2}$ region of Arp\,220.
Each panel shows the polarization vectors (black lines) satisfying $PI/\sigma_{PI} \ge4$ per channel overlaid on the polarized intensity (colormap) and total intensity of the blueshifted (blue contours) and redshifted (red contours) outflows. The velocity, range of polarization fraction, and polarization angles are noted at the top of each panel. Dust continuum image contours, as shown in Figure~\ref{fig:fig2}, are also presented. Black crosses mark the central positions of Arp\,220\,E and Arp\,220\,W.
 \label{App:fig2}}
\end{figure*}

Figure~\ref{App:fig2} shows the polarization measurements within the velocity channels of the outflows that satisfy the condition $PI/\sigma_{\rm{PI}}\ge4$.

\begin{figure*}[ht!]
\includegraphics[width=\textwidth]{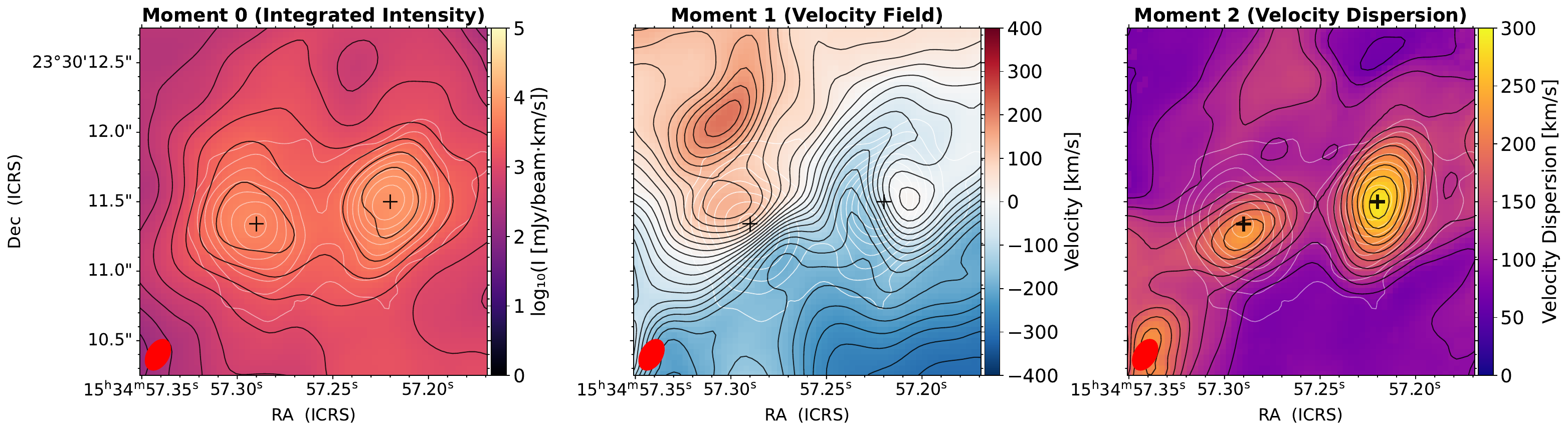}
\caption{Moments of the CO(3-2) line emission within the central $1\times1$\,kpc$^{2}$ region of Arp\,220. The moments 0 (left), 1 (middle), and 2 (right) maps are shown with contours (black lines) every $0.2$~dex, $50$\,km/s, and $20$\,km/s, respectively. The dust continuum image contours (white) from Figure~\ref{fig:fig2} are also displayed. Black crosses mark the central positions of Arp\,220\,E and Arp\,220\,W.
 \label{App:fig1}}
\end{figure*}

Figure~\ref{App:fig1} shows the moments of the CO(3-2) emission line within the central $1\times1$ kpc$^{2}$ region of Arp\,220.

%%%%%%%%%%%%%%

%\vfil\eject
%%%%%%%%%%%%%%

%%%%%%%%%%%%%%
%%%%%%%%%%%%%%

%% For this sample we use BibTeX plus aasjournalv7.bst to generate the

%\bibliography{references}{}
%\bibliographystyle{aasjournalv7}



%\begin{appendix}

\vfil\eject

\end{document}